\documentclass[seceqn]{elsart}
\usepackage{graphicx}
\newcommand{\order}[1]{\mathcal{O}\left( #1 \right)}
\newcommand{\ep}{\varepsilon}
\newcommand{\Li}[1]{\mathop{\mathrm{Li}}\nolimits_{#1}}

\begin{document}

\begin{flushright}
\newlength{\prepw}
\settowidth{\prepw}{Alberta Thy 00-00}
\begin{minipage}{\prepw}
TTP09-08\\
SFB/CPP-09-29\\
Alberta Thy 05-09
\end{minipage}
\end{flushright}

\begin{frontmatter}
\title{Three-loop on-shell Feynman integrals with two masses}
\author[Karlsruhe]{S.~Bekavac},
\author[Novosibirsk,Karlsruhe]{A.G.~Grozin},
\author[Edmonton]{D.~Seidel} and
\author[Moscow]{V.A.~Smirnov}
\address[Karlsruhe]{Institut f\"ur Theoretische Teilchenphysik,
Universit\"at Karlsruhe, 76128 Karlsruhe, Germany}
\address[Novosibirsk]{Budker Institute of Nuclear Physics,
Novosibirsk 630090, Russia}
\address[Edmonton]{Department of Physics,
University of Alberta, Edmonton, AB, Canada T6G 2J1}
\address[Moscow]{Nuclear Physics Institute, Moscow State University,
Moscow, Russia}
\begin{abstract}
All three-loop on-shell QCD Feynman integrals with two masses
can be reduced to 27 master integrals.
Here we calculate these master integrals, expanded in $\ep$,
both exactly in the mass ratio and as series in limiting cases.
\end{abstract}
\end{frontmatter}

\section{Introduction}
\label{S:Intro}

\begin{figure}[ht]
\begin{center}
\begin{picture}(42,17)
\put(21,8.5){\makebox(0,0){\includegraphics{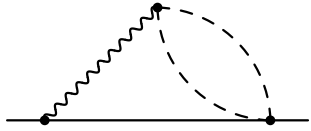}}}
\end{picture}
\end{center}
\caption{The generic class of two-loop on-shell diagrams with two masses.
Solid lines have mass $M$, dashed lines mass $m$, and wavy lines
are massless.}
\label{F:T2}
\end{figure}

\begin{figure}[ht]
\begin{center}
\begin{picture}(103,25)
\put(6,14){\makebox(0,0){\includegraphics{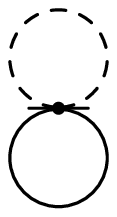}}}
\put(23,14){\makebox(0,0){\includegraphics{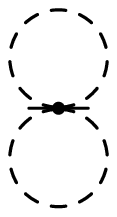}}}
\put(50,14){\makebox(0,0){\includegraphics{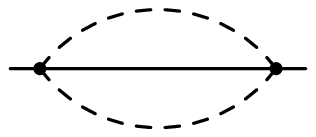}}}
\put(87,14){\makebox(0,0){\includegraphics{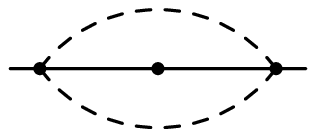}}}
\put(6,0){\makebox(0,0)[b]{2.1}}
\put(23,0){\makebox(0,0)[b]{2.2}}
\put(50,4){\makebox(0,0)[b]{3.1}}
\put(87,4){\makebox(0,0)[b]{3.1a}}
\end{picture}
\end{center}
\caption{The two-loop master integrals.}
\label{F:M2}
\end{figure}

Massive on-shell Feynman integrals have numerous applications.
We consider two-leg diagrams, where the external particle
has mass $M$ and an on-shell momentum $p$ ($p^2=M^2$), in QCD or QED%
\footnote{These results are also useful in the electroweak theory,
or more general field theories;
but there additional classes of diagrams appear.}.
On-shell Feynman integrals with a single mass $M$
have been investigated
at two~\cite{GBS,Broadhurst:1991fi,FT}
and three~\cite{Laporta:1996mq,Melnikov:2000zc,Marquard:2007uj} loops.
Starting from two loops, diagrams with loops of massive particles
having a different mass $m$ appear.
Such diagrams are non-trivial functions of the ratio
\begin{equation}
x = \frac{m}{M}\,.
\label{Intro:x}
\end{equation}
At two loops, there is one generic class of diagram%
\footnote{We use the following definitions:
a \emph{class} consists of diagrams with identical denominators,
or whose denominators can be made identical
by linear substitutions for their integration momenta;
a class is \emph{generic} if it contains a maximum number
of lines (denominators);
contracting some line(s) in a class,
one obtains \emph{contracted} classes.}
in QCD and QED (Fig.~\ref{F:T2}).
The corresponding Feynman integrals can be reduced
to 4 master integrals (Fig.~\ref{F:M2}),
coefficients being rational functions of $d=4-2\ep$ and $x$;
a reduction algorithm has been constructed
in Ref.~\cite{Davydychev:1998si}%
\footnote{In this paper, another integral was used as master
instead of the last one in Fig.~\ref{F:M2}.
It is, of course, easy to reduce this last integral
to the master ones of Ref.~\cite{Davydychev:1998si},
solve for this master integral, and then express all results
via the integrals of Fig.~\ref{F:M2}.}.

\begin{figure}[ht]
\begin{center}
\begin{picture}(151,24.5)
\put(21,14.75){\makebox(0,0){\includegraphics{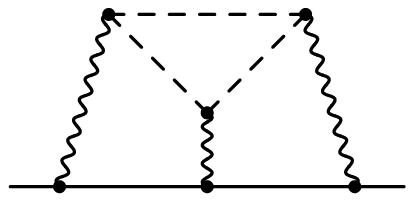}}}
\put(68,13.5){\makebox(0,0){\includegraphics{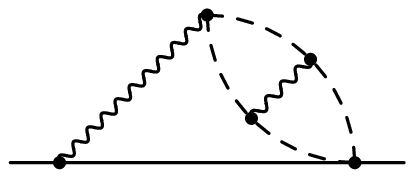}}}
\put(120,11){\makebox(0,0){\includegraphics{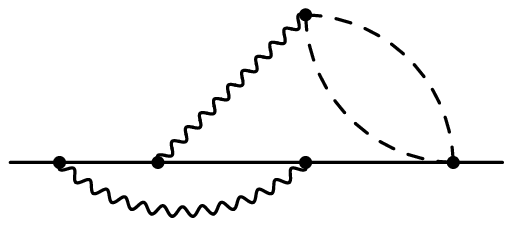}}}
\end{picture}
\end{center}
\caption{The generic classes of three-loop on-shell diagrams with two masses.}
\label{F:T3}
\end{figure}

\begin{figure}[ht]
\begin{center}
\begin{picture}(136,33)
\put(21,17){\makebox(0,0){\includegraphics{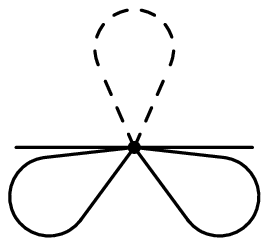}}}
\put(21,1.5){\makebox(0,0){3.1}}
\put(68,17){\makebox(0,0){\includegraphics{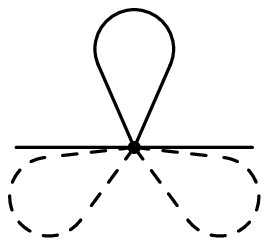}}}
\put(68,1.5){\makebox(0,0){3.2}}
\put(115,17){\makebox(0,0){\includegraphics{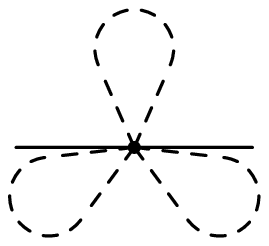}}}
\put(115,1.5){\makebox(0,0){3.3}}
\end{picture}
\end{center}
\caption{The master integrals with 3 lines.}
\label{F:M3}
\end{figure}

\begin{figure}[ht]
\begin{center}
\begin{picture}(103,25)
\put(11,14){\makebox(0,0){\includegraphics{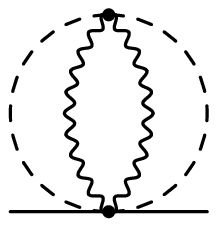}}}
\put(11,1.5){\makebox(0,0){4.1}}
\put(38,14){\makebox(0,0){\includegraphics{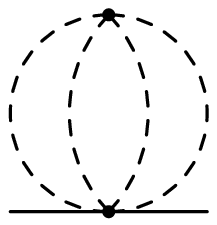}}}
\put(38,1.5){\makebox(0,0){4.2}}
\put(65,14){\makebox(0,0){\includegraphics{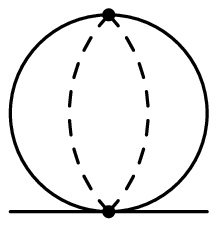}}}
\put(65,1.5){\makebox(0,0){4.3}}
\put(92,14){\makebox(0,0){\includegraphics{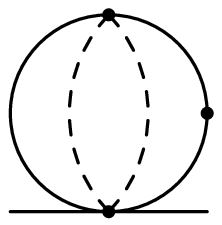}}}
\put(92,1.5){\makebox(0,0){4.3a}}
\end{picture}
\end{center}
\caption{The master integrals with 4 lines: vacuum bubbles.}
\label{F:M4v}
\end{figure}

\begin{figure}[ht]
\begin{center}
\begin{picture}(89,73)
\put(21,12){\makebox(0,0){\includegraphics{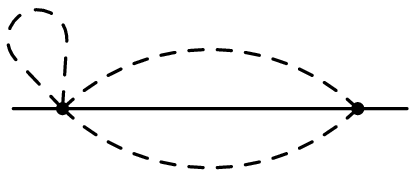}}}
\put(21,1.5){\makebox(0,0){4.7}}
\put(68,12){\makebox(0,0){\includegraphics{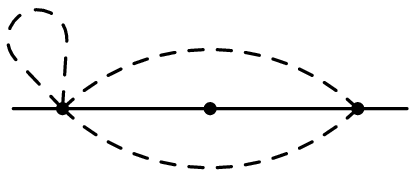}}}
\put(68,1.5){\makebox(0,0){4.7a}}
\put(21,38){\makebox(0,0){\includegraphics{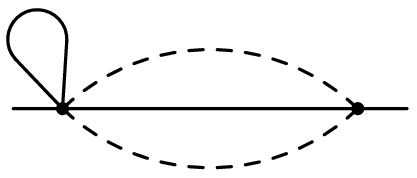}}}
\put(21,27.5){\makebox(0,0){4.6}}
\put(68,38){\makebox(0,0){\includegraphics{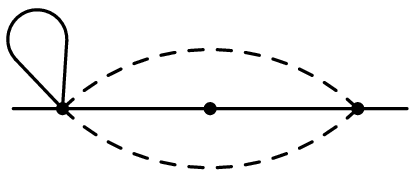}}}
\put(68,27.5){\makebox(0,0){4.6a}}
\put(21,64){\makebox(0,0){\includegraphics{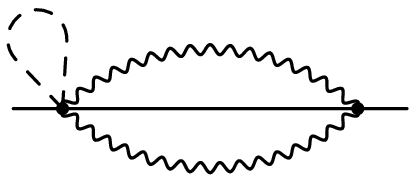}}}
\put(21,53.5){\makebox(0,0){4.4}}
\put(68,64){\makebox(0,0){\includegraphics{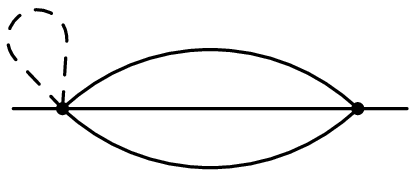}}}
\put(68,53.5){\makebox(0,0){4.5}}
\end{picture}
\end{center}
\caption{The master integrals with 4 lines: two-loop sunsets.}
\label{F:M4s2}
\end{figure}

\begin{figure}[ht]
\begin{center}
\begin{picture}(136,23)
\put(21,13){\makebox(0,0){\includegraphics{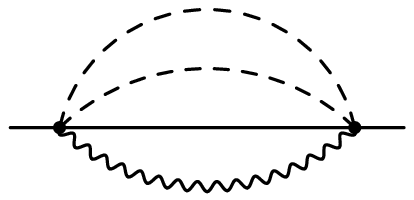}}}
\put(21,1.5){\makebox(0,0){4.8}}
\put(68,13){\makebox(0,0){\includegraphics{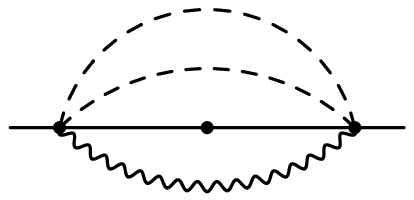}}}
\put(68,1.5){\makebox(0,0){4.8a}}
\put(115,13){\makebox(0,0){\includegraphics{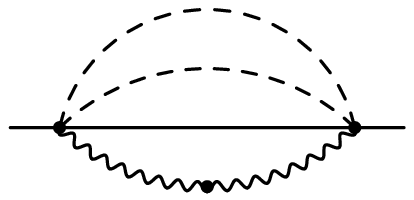}}}
\put(115,1.5){\makebox(0,0){4.8b}}
\end{picture}
\end{center}
\caption{The master integrals with 4 lines: three-loop sunsets.}
\label{F:M4s3}
\end{figure}

\begin{figure}[ht]
\begin{center}
\begin{picture}(99,90.5)
\put(21,9.25){\makebox(0,0){\includegraphics{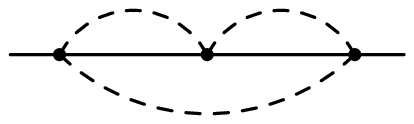}}}
\put(21,1.5){\makebox(0,0){5.4}}
\put(68,9.25){\makebox(0,0){\includegraphics{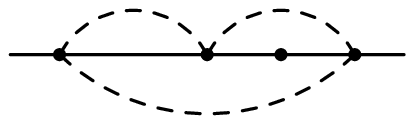}}}
\put(68,1.5){\makebox(0,0){5.4a}}
\put(21,32){\makebox(0,0){\includegraphics{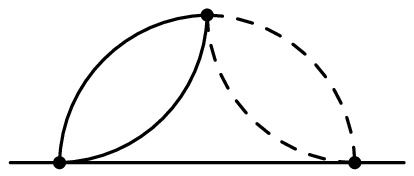}}}
\put(21,22){\makebox(0,0){5.3}}
\put(68,32){\makebox(0,0){\includegraphics{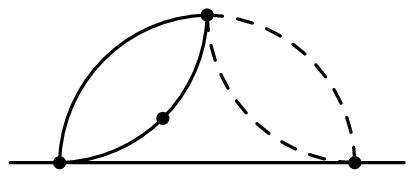}}}
\put(68,22){\makebox(0,0){5.3a}}
\put(21,57){\makebox(0,0){\includegraphics{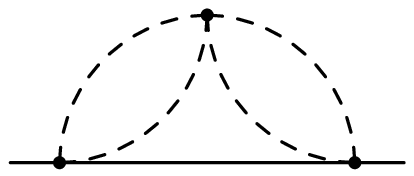}}}
\put(21,47){\makebox(0,0){5.2}}
\put(68,57){\makebox(0,0){\includegraphics{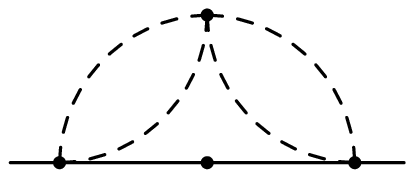}}}
\put(68,47){\makebox(0,0){5.2a}}
\put(21,82){\makebox(0,0){\includegraphics{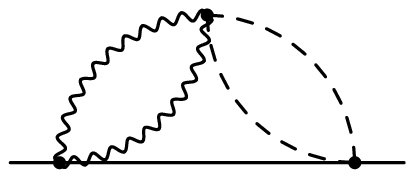}}}
\put(21,72){\makebox(0,0){5.1}}
\put(68,82){\makebox(0,0){\includegraphics{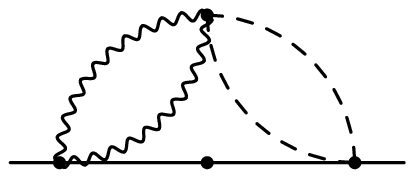}}}
\put(68,72){\makebox(0,0){5.1a}}
\end{picture}
\end{center}
\caption{The master integrals with 5 lines.}
\label{F:M5}
\end{figure}

\begin{figure}[ht]
\begin{center}
\begin{picture}(136,20)
\put(21,11.5){\makebox(0,0){\includegraphics{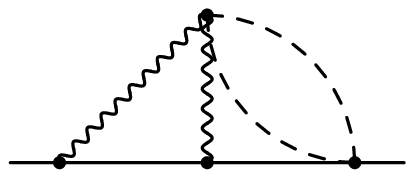}}}
\put(21,1.5){\makebox(0,0){6.1}}
\put(68,11.5){\makebox(0,0){\includegraphics{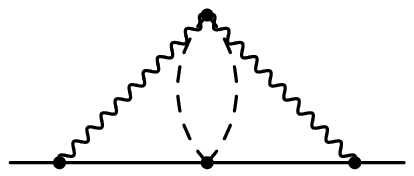}}}
\put(68,1.5){\makebox(0,0){6.2}}
\put(115,11.5){\makebox(0,0){\includegraphics{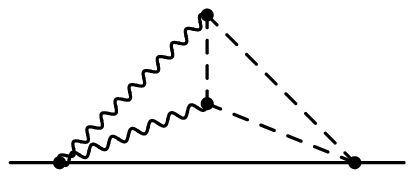}}}
\put(115,1.5){\makebox(0,0){6.3}}
\end{picture}
\end{center}
\caption{The master integrals with 6 lines.}
\label{F:M6}
\end{figure}

The first three-loop on-shell calculation with two different masses
has been done by Laporta and Remiddi~\cite{Laporta:1992pa}.
At three loops, there are three generic classes of diagrams
in QCD (Fig.~\ref{F:T3}).
Recently, three of us together with M.~Steinhauser
have demonstrated~\cite{Bekavac:2007tk}
that the corresponding Feynman integrals can be reduced
to 27 master integrals (Figs.~\ref{F:M3}--\ref{F:M6})
(see also~\cite{Bekavac:2007hd}).
The reduction is performed by the \texttt{C++}
program \texttt{Crusher}~\cite{Crusher} which implements
the Laporta algorithm~\cite{Laporta:2001dd} to solve integration by
parts identities~\cite{Chetyrkin:1981qh}.
This is done exactly at general $d$;
simple polynomial operations are done
by \texttt{GiNaC}~\cite{Bauer:2000cp},
and simplification of rational expressions in $d$ and $x$
by \texttt{Fermat}~\cite{Fermat}.
On-shell mass and wave-function renormalization of a heavy quark
(e.g., $b$) in QCD with another massive flavour (e.g., $c$)
have been calculated at three loops.
In the present paper, we present details of the calculation
of the master integrals used in~\cite{Bekavac:2007tk}.
These integrals can be used in many other
three-loop on-shell calculations.

Our results can also be used for another important kind of diagrams,
namely, diagrams with two different non-zero masses and
any number of external lines with small momenta.
If we expand in them, these diagrams reduce
to vacuum integrals with two masses.
And such integrals are particular cases of on-shell integrals
considered in the present paper
(the incoming leg and the outgoing one are attached
to one and the same vertex).
Such diagrams reduce to the master integrals
in Figs.~\ref{F:M3}, \ref{F:M4v}.
Therefore, the master integrals we are investigating here
are also useful for another wide area of applications.

We shall use the Minkowski notation.
All denominators will contain $-i0$.
Loop integrals will be divided by $i\pi^{d/2}$
for each loop.
For example,
\begin{eqnarray}
\raisebox{-7mm}{\begin{picture}(10,13)
\put(5,8){\makebox(0,0){\includegraphics{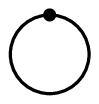}}}
\put(5,2){\makebox(0,0){$a$}}\end{picture}}
&=& \frac{1}{i\pi^{d/2}}
\int \frac{d^d k}{(M^2-k^2-i0)^a}
= \frac{\Gamma(a-d/2)}{\Gamma(a)} M^{d-2a}\,,
\label{Intro:IM}\\
\raisebox{-7mm}{\begin{picture}(10,13)
\put(5,8){\makebox(0,0){\includegraphics{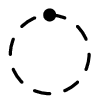}}}
\put(5,2){\makebox(0,0){$a$}}\end{picture}}
&=& \frac{\Gamma(a-d/2)}{\Gamma(a)} m^{d-2a}\,.
\label{Intro:Im}
\end{eqnarray}
Powers of denominators are either written near the corresponding lines,
or indicated by dots (if they are some specific small integers).
The master integrals with 3 lines (Fig.~\ref{F:M3})
are products of~(\ref{Intro:IM}), (\ref{Intro:Im}).

\section{Two-loop sunset diagrams}
\label{S:S2}

\begin{figure}[ht]
\begin{center}
\begin{picture}(106,17)
\put(16,10){\makebox(0,0){\includegraphics{s32.eps}}}
\put(53,10){\makebox(0,0){\includegraphics{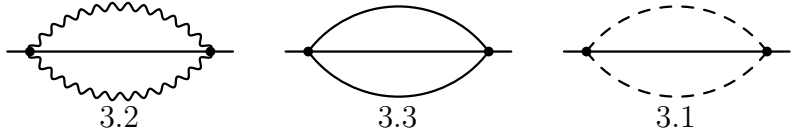}}}
\put(90,10){\makebox(0,0){\includegraphics{s31.eps}}}
\put(16,0){\makebox(0,0)[b]{3.2}}
\put(53,0){\makebox(0,0)[b]{3.3}}
\put(90,0){\makebox(0,0)[b]{3.1}}
\end{picture}
\end{center}
\caption{Two-loop sunset diagrams.}
\label{F:S2}
\end{figure}

The master integrals 4.4--4.7a (Fig.~\ref{F:M4s2})
are products of two-loop sunset diagrams (Fig.~\ref{F:S2})
and one-loop vacuum bubbles.
The first sunset diagram can be expressed via $\Gamma$ functions
for any powers of denominators:
\begin{eqnarray}
&&\raisebox{-9mm}{\begin{picture}(32,20)
\put(16,10){\makebox(0,0){\includegraphics{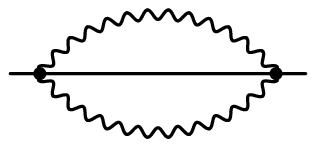}}}
\put(16,18){\makebox(0,0){$a_1$}}
\put(16,2){\makebox(0,0){$a_2$}}
\put(16,8){\makebox(0,0){$a_3$}}\end{picture}}
= \frac{\Gamma(d/2-a_1) \Gamma(d/2-a_2) \Gamma(a_1+a_2-d/2)}%
{\Gamma(a_1) \Gamma(a_2) \Gamma(a_3)}
\nonumber\\
&&{}\times\frac{\Gamma(a_1+a_2+a_3-d) \Gamma(2(d-a_1-a_2)-a_3)}%
{\Gamma(d-a_1-a_2) \Gamma(3d/2-a_1-a_2-a_3)}
M^{3d-2(a_1+a_2+a_3)}\,.
\label{S2:1}
\end{eqnarray}

The non-trivial single-scale master integral (3.3 in Fig.~\ref{F:S2})
was investigated in Ref.~\cite{Broadhurst:1991fi}.
It can be written as
\begin{eqnarray}
&&\raisebox{-6.5mm}{\includegraphics{s33.eps}} =
- \frac{\Gamma^2(1+\ep)}{(1-2\ep)(1-3\ep)(2-3\ep)}
\nonumber\\
&&{}\times\Biggl[
3 \frac{1-6\ep}{\ep^2} + \frac{7}{\ep}
\frac{\Gamma(1-\ep) \Gamma^2(1+2\ep) \Gamma(1+3\ep)}%
{\Gamma^2(1+\ep) \Gamma(1+4\ep)}
\nonumber\\
&&\hphantom{{}\times\Biggl[\Biggr.} + \frac{8}{3} \ep \pi^2 2^{-6\ep}
\frac{\Gamma(1+2\ep) \Gamma(1+3\ep)}{\Gamma^5(1+\ep)}
+ 24 \ep^3 B_4(\ep) \Biggr]
M^{2-4\ep}\,,
\label{S2:2}
\end{eqnarray}
where
\begin{eqnarray}
B_4(\ep) &=& \frac{8}{3 \ep^2 (1+2\ep)}
{}_3\!F_2 \left( \left.
\begin{array}{c}
1, \frac{1}{2}-\ep, \frac{1}{2}-\ep\\
\frac{3}{2}, \frac{3}{2}+\ep
\end{array}
\right| 1 \right)
\nonumber\\
&&{} - \frac{7}{24\ep^4} \left(
\frac{\Gamma(1-\ep) \Gamma^2(1+2\ep) \Gamma(1+3\ep)}%
{\Gamma^2(1+\ep) \Gamma(1+4\ep)}
- 1 \right)
\nonumber\\
&&{} - \frac{\pi^2 2^{-6\ep}}{3\ep^2}
\frac{\Gamma(1+2\ep) \Gamma(1+3\ep)}{\Gamma^5(1+\ep)}
\nonumber\\
&=& 16 \Li4\left(\frac{1}{2}\right)
+ \frac{2}{3} \log^2 2 \left( \log^2 2 - \pi^2 \right)
- \frac{13}{180} \pi^4
+ \order{\ep}\,.
\label{S2:B4}
\end{eqnarray}
The expansion of $B_4(\ep)$ up to $\order{\ep^3}$
can be found in~\cite{Broadhurst:1996az}.
The hypergeometric function can also easily be expanded in $\ep$
with the help of the package \texttt{HypExp}~\cite{HypExp}.

The two master sunset integrals with two masses (Fig.~\ref{F:M2})
have been calculated up to $\order{\ep^0}$ in Ref.~\cite{Berends:1997vk},
and up to $\order{\ep^5}$ in Ref.~\cite{Argeri:2002wz}%
\footnote{they were also calculated up to $\order{\ep}$
as series in $x$ up to $x^6$~\cite{Onishchenko:2002ri}.}.
Here we obtain exact results for them to $\order{\ep^3}$.
To this end we use the method of differential equations~\cite{Kotikov:1990kg}.
The class of two-loop sunset diagrams with two masses has two master integrals.
This leads to a system of coupled differential equations.
The first one is
\begin{eqnarray}
\frac{d}{dx}
\raisebox{-6.5mm}{\includegraphics{s31.eps}} &=&
2 \left(\frac{1}{x+1}+\frac{1}{x-1}\right)
\raisebox{-6.5mm}{\includegraphics{s31.eps}}
\nonumber\\
&&{} - 32 \ep M^4 \left(\frac{1}{x+1} + \frac{1}{x-1}\right)
\raisebox{-6.5mm}{\includegraphics{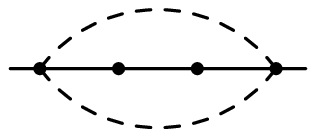}}
\nonumber\\
&&{} - \frac{2}{M^2}
\left(\frac{2}{x} - \frac{1}{x+1} - \frac{1}{x-1}\right)
\raisebox{-10.5mm}{\includegraphics{s21.eps}}
\nonumber\\
&&{} + \frac{1}{M^2}
\left(\frac{2}{x} - \frac{1}{x+1} - \frac{1}{x-1}\right)
\raisebox{-10.5mm}{\includegraphics{s22.eps}}\,,
\label{eq::deq2l}
\end{eqnarray}
where the coefficients have already been expanded in $\ep$
and only the leading term in $\ep$ for each coefficient is written.
As usual the integrals of contracted classes are assumed to be known
and so are part of the inhomogeneous term of the differential equation.
If the system is solved order by order in $\ep$
the integral with two dots (3.1b) decouples from~(\ref{eq::deq2l}).
The equation for this integral is
\begin{equation}
\frac{d}{dx} \raisebox{-6.5mm}{\includegraphics{s31b.eps}} =
\frac{1}{x} \raisebox{-6.5mm}{\includegraphics{s31b.eps}}\,.
\end{equation}
Note that here $\ep$ is set to zero in all coefficients.
Of course all integrals which appear in~(\ref{eq::deq2l})
are also present in this equation,
but they only contribute at higher orders in $\ep$.
Looking at the structure of the coefficients in this system
one can see that it is possible to get solutions for the master integrals
in terms of harmonic polylogarithms~\cite{Remiddi:1999ew}
(this is also true if one writes the coefficients
to arbitrary order in $\ep$).
We have used the package \texttt{HPL}~\cite{HPL}
to implement the integration of the differential equations.

After integrating there is one constant per order in $\ep$
and per integral which is not determined.
In the case of the integral 3.1 (Fig.~\ref{F:S2})
one constant can be determined in the limit $x\to 0$,
where this integral reduces to 3.2.
For the integral 3.1b the limit $x\to 0$ gives the correct result
independently of the choice of the constant,
as the constant in the integral is proportional to $x$
and drops out in this limit.
To get this constant we use the limit $x\to 1$,
where the resulting integral can be reduced to 3.3.

The result for the master integral 3.1 (Fig.~\ref{F:S2})
up to the finite part in $\ep$ reads
\begin{eqnarray}
&&{}\hspace{-7mm}\raisebox{-6.5mm}{\includegraphics{s31.eps}}
= M^{2-4\ep} \Gamma^2(1+\ep)
\Biggl[ - \frac{1}{\ep^2} \left(x^2 + \frac{1}{2}\right)
\nonumber\\
&&{}\hspace{-7mm}{} + \frac{1}{\ep} \left(4x^2\,H(0;x) - 3 x^2 - \frac{5}{4}\right)
+ 2 (x^2-1)^2 \left(H(-1,0;x) - H(1,0;x) - \frac{\pi^2}{6}\right)
\nonumber\\
&&{}\hspace{-7mm}{} - 4 x^2 (x^2+2) H(0,0;x) + 14 x^2\,H(0;x)
- 6 x^2 - \frac{11}{8}
+ \order{\ep} \Biggr]\,,
\label{S2:res}
\end{eqnarray}
where $H$ denote the harmonic polylogarithms.
Up to this order they can of course be expressed
in terms of logarithms and ordinary polylogarithms,
but this gets impossible in higher orders in $\ep$.
We have calculated the diagram 3.1a (Fig.~\ref{F:S2})
(which can be reduced to 3.1b) up to order $\ep^3$,
but we refrain from presenting the result in a written form in this paper.
Instead we refer to the web-site~\cite{site},
where all our results can be found in the form
of a \texttt{Mathematica} package.
Our results agree with~\cite{Argeri:2002wz}.

We also used another approach.
The Mellin--Barnes representation
of the massive one-loop self-energy is~\cite{Boos:1990rg}%
\footnote{For a simpler derivation, see~\cite{books}.}
\begin{eqnarray}
&&\raisebox{-8mm}{\begin{picture}(22,18)
\put(11,9){\makebox(0,0){\includegraphics{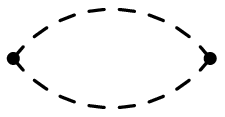}}}
\put(11,16){\makebox(0,0){$a_1$}}
\put(11,2){\makebox(0,0){$a_2$}}
\end{picture}}
= \frac{1}{i\pi^{d/2}} \int
\frac{d^d k}{\left[m^2-k^2-i0\right]^{a_1}
\left[m^2-(k+p)^2-i0\right]^{a_2}}
\nonumber\\
&&{} = \frac{m^{d-2(a_1+a_2)}}{\Gamma(a_1) \Gamma(a_2)}
\frac{1}{2\pi i} \int_{-i\infty}^{+i\infty} dz\,\Gamma(-z)
\nonumber\\
&&{}\times\frac{\Gamma(a_1+z) \Gamma(a_2+z)
\Gamma(a_1+a_2-d/2+z)}{\Gamma(a_1+a_2+2z)} m^{-2z}
\raisebox{-3mm}{\begin{picture}(22,8)
\put(11,4){\makebox(0,0){\includegraphics{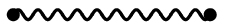}}}
\put(11,6){\makebox(0,0){$-z$}}
\end{picture}}\,,
\label{S2:MB}
\end{eqnarray}
where the integration contour is chosen in such a way
that poles of $\Gamma$ functions with $+z$ are to the left of it
and of those with $-z$ are to the right.
We obtain
\begin{eqnarray}
&&\raisebox{-9mm}{\begin{picture}(32,20)
\put(16,10){\makebox(0,0){\includegraphics{s31.eps}}}
\put(16,18){\makebox(0,0){$a_1$}}
\put(16,2){\makebox(0,0){$a_2$}}
\put(16,8){\makebox(0,0){$a_3$}}
\end{picture}} =
\frac{M^{d-2a_3} m^{d-2(a_1+a_2)}}%
{\Gamma(a_1) \Gamma(a_2) \Gamma(a_3)}
\frac{1}{2\pi i} \int_{-i\infty}^{+i\infty} dz\,x^{-2z}
\nonumber\\
&&{}\times \Gamma(-z) \Gamma(a_3-d/2-z)
\nonumber\\
&&{}\times
\frac{\Gamma(a_1+z) \Gamma(a_2+z) \Gamma(a_1+a_2-d/2+z)
\Gamma(d-a_3+2z)}{\Gamma(d-a_3+z) \Gamma(a_1+a_2+2z)}\,.
\label{S2:MB1}
\end{eqnarray}
This result is a particular case of~\cite{Davydychev:1998si}.

\section{Three-loop vacuum bubbles}
\label{S:V3}

The diagrams in the first row of Fig.~\ref{F:M4v}
are vacuum bubbles.
The first one, 4.1, can be expressed via $\Gamma$ functions
for any powers of denominators:
\begin{eqnarray}
&&\raisebox{-10mm}{\begin{picture}(28,22)
\put(14,11){\makebox(0,0){\includegraphics{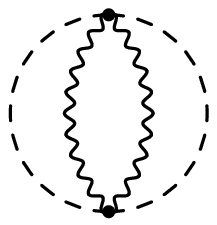}}}
\put(8,11){\makebox(0,0){$a_1$}}
\put(20,11){\makebox(0,0){$a_2$}}
\put(2,11){\makebox(0,0){$a_3$}}
\put(26,11){\makebox(0,0){$a_4$}}
\end{picture}} =
\frac{\Gamma(d/2-a_1) \Gamma(d/2-a_2) \Gamma(a_1+a_2-d/2)}%
{\Gamma(a_1) \Gamma(a_2) \Gamma(a_3) \Gamma(a_4) \Gamma(d/2)}
\nonumber\\
&&{}\times \Gamma(a_1+a_2+a_3-d) \Gamma(a_1+a_2+a_4-d)
\nonumber\\
&&{}\times
\frac{\Gamma(a_1+a_2+a_3+a_4-3d/2)}{\Gamma(2(a_1+a_2-d)+a_3+a_4)}\,
m^{3d-2(a_1+a_2+a_3+a_4)}\,.
\label{V3:1}
\end{eqnarray}

The non-trivial single-scale master integral (4.2 in Fig.~\ref{F:M4v})
was investigated in Ref.~\cite{Broadhurst:1991fi}:
\begin{eqnarray}
&&\raisebox{-10mm}{\includegraphics{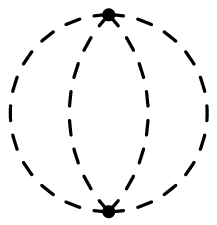}} =
\frac{4\Gamma^3(1+\ep)}{3(1-\ep)(1-2\ep)(1-3\ep)(2-3\ep)}
\Biggl[ 3 \frac{1-6\ep}{\ep^3}
\nonumber\\
&&{} + \frac{7}{\ep^2}
\frac{\Gamma(1-\ep) \Gamma^2(1+2\ep) \Gamma(1+3\ep)}%
{\Gamma^2(1+\ep) \Gamma(1+4\ep)}
+ 24 \ep^2 B_4(\ep) \Biggr]
m^{4-6\ep}\,,
\label{V3:2}
\end{eqnarray}
where $B_4(\ep)$ is given by Eq.~(\ref{S2:B4}).

As in the case of the two-loop sunset diagrams,
the vacuum bubble diagrams 4.3 and 4.3a can be calculated
by solving the corresponding differential equations
in terms of harmonic polylogarithms.
We have used the solution of diagram 4.1 (see Eq.~(\ref{V3:1}))
as initial condition in the limit $x\to0$.
Taking the limit $x\to 1$ we recover the expanded version of Eq.~(\ref{V3:2}).
For example,
\begin{eqnarray}&&\raisebox{-10mm}{\includegraphics{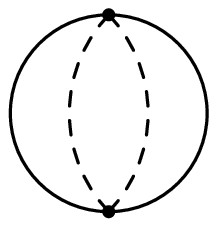}} =
M^{4-6\ep} \Gamma^3(1+\ep) \Biggl[
\frac{x^4 + 4 x^2 + 1}{3 \ep^3}
\nonumber\\
&&{} - \frac{1}{\ep^2}
\left( 2 x^2 (x^2 + 2) H(0;x)
- \frac{1}{6} (7 x^4 + 32 x^2 + 7) \right)
\nonumber\\
&&{} + \frac{1}{\ep} \biggl( 4 x^2 (3 x^2 + 2) H(0,0;x)
- x^2 (7 x^2 + 16) H(0;x)
\nonumber\\
&&\qquad{} + \frac{5}{12} (5 x^4 + 32 x^2 + 5) \biggr)
+ \order{1} \Biggr]\,.
\label{V3:3}
\end{eqnarray}
The analytic results expanded up to $\order{\ep^2}$
of all vacuum-diagrams depicted in Fig.~(\ref{F:M4v})
can be obtained on the aforementioned webpage.

For the vacuum bubble with two masses we obtain,
using the Mellin--Barnes representation~(\ref{S2:MB}),
\begin{eqnarray}
&&\raisebox{-10mm}{\begin{picture}(28,22)
\put(14,11){\makebox(0,0){\includegraphics{v3.eps}}}
\put(8,11){\makebox(0,0){$a_1$}}
\put(20,11){\makebox(0,0){$a_2$}}
\put(2,11){\makebox(0,0){$a_3$}}
\put(26,11){\makebox(0,0){$a_4$}}
\end{picture}} =
\frac{M^{2(d-a_3-a_4)} m^{d-2(a_1+a_2)}}%
{\Gamma(a_1) \Gamma(a_2) \Gamma(a_3) \Gamma(a_4) \Gamma(d/2)}
\frac{1}{2\pi i} \int_{-i\infty}^{+i\infty} dz\,x^{-2z}
\nonumber\\
&&{}\times
\frac{\Gamma(-z) \Gamma(a_3-d/2-z) \Gamma(a_4-d/2-z)
\Gamma(a_3+a_4-d-z)}{\Gamma(a_3+a_4-d-2z)}
\nonumber\\
&&{}\times
\frac{\Gamma(d/2+z) \Gamma(a_1+z) \Gamma(a_2+z)
\Gamma(a_1+a_2-d/2+z)}{\Gamma(a_1+a_2+2z)}
\label{V3:MB}
\end{eqnarray}
(this is a particular case of a more general result
recently derived in~\cite{Bytev:2009mn}).
This diagram is, of course, symmetric with respect to
$a_{1,2}\leftrightarrow a_{3,4}$, $M\leftrightarrow m$,
though this is not evident in~(\ref{V3:MB}).

\section{Three-loop sunset diagrams}
\label{S:S3}

Here we consider three-loop on-shell sunsets with masses $M$, $m$, $m$, 0
(Fig.~\ref{F:M4s3}).
The first of them has been calculated up to $\order{\ep^3}$
in Ref.~\cite{Mastrolia:2002tv}, though only the $\order{\ep^1}$ term
is presented in the paper.

Doing a naive reduction to master integrals
of all three-loop on-shell sunsets with masses $M$, $m$, $m$, 0
one finds four ``master integrals''.
As already noticed in~\cite{Mastrolia:2002tv},
one of these integrals decouples completely
when setting up a higher order differential equation
for the integral 4.8 (Fig.~\ref{F:M4s3}).
We have performed a reduction for all classes of the
different master integrals presented in this paper.
It turns out that in the course of the reduction of the class
to which the integral 5.1 (Fig.~\ref{F:M5}) belongs,
an equation is generated which contains no five-line master integrals
and connects the integral 4.1 (Fig.~\ref{F:M4v})
and one of the integrals from the class discussed in this section.
So it gets manifest that one of the four integrals
found by the naive reduction is reducible with integration by part identities.

For example, for 4.8 we obtain
\begin{eqnarray}
&&\raisebox{-6.5mm}{\includegraphics{j48.eps}} =
M^{4-6\ep} \Gamma^3(1+\ep) \Biggl[
\frac{x^2 (x^2+2)}{3 \ep^3}
\nonumber\\
&&{} + \frac{1}{\ep^2} \left(
- 2 x^2 (x^2 + 1) H(0;x)
+ \frac{1}{6} (7 x^4 + 15 x^2 - 1) \right)
\nonumber\\
&&{} + \frac{1}{\ep} \biggl(
4 x^2 (3 x^2 + 1) H(0,0;x) - 7 x^2 (x^2 + 1) H(0;x)
\nonumber\\
&&\qquad{} + \frac{1}{36} (75 x^4 + 213 x^2 - 35) \biggr)
+ \order{1} \Biggr]\,.
\label{S3:1}
\end{eqnarray}
The analytic results for all three master integrals
in Fig.~\ref{F:M4s3} up to order $\ep$
are presented on the aforementioned webpage.

Using the Mellin--Barnes representation~(\ref{S2:MB}),
we obtain
\begin{eqnarray}
&&\raisebox{-9mm}{\begin{picture}(42,26)
\put(21,13){\makebox(0,0){\includegraphics{j48.eps}}}
\put(21,24){\makebox(0,0){$a_1$}}
\put(21,18){\makebox(0,0){$a_2$}}
\put(21,2){\makebox(0,0){$a_3$}}
\put(21,8){\makebox(0,0){$a_4$}}
\end{picture}} =
\frac{M^{2(d-a_3-a_4)} m^{d-2(a_1+a_2)} \Gamma(d/2-a_3)}%
{\Gamma(a_1) \Gamma(a_2) \Gamma(a_3) \Gamma(a_4)}
\nonumber\\
&&{}\times \frac{1}{2\pi i} \int_{-i\infty}^{+i\infty} dz\,x^{-2z}
\Gamma(a_3-d/2-z) \Gamma(a_3+a_4-d-z) \Gamma(d/2+z)
\nonumber\\
&&{}\times
\frac{\Gamma(a_1+z) \Gamma(a_2+z)
\Gamma(a_1+a_2-d/2+z) \Gamma(2(d-a_3)-a_4+2z)}%
{\Gamma(d-a_3+z) \Gamma(3d/2-a_3-a_4+z) \Gamma(a_1+a_2+2z)}\,.
\label{S3:MB}
\end{eqnarray}

\section{Master integrals with 5 lines}
\label{S:M5}

We have calculated the master integrals 5.1 and 5.1a (Fig.~\ref{F:M5})
analytically up to $\order{\ep}$ with the differential equations method.
The integrals 5.4 and 5.4a are obtained with the same method
up to $\order{\ep^2}$.
The integrals 5.2, 5.2a, 5.3, 5.3a are obtained up to $\order{\ep^{-1}}$
with that method.

For the two classes 5.1 and 5.4
the differential equations have a structure
with which it is possible to integrate them
with the help of harmonic polylogarithms.
So in principle, if the initial conditions are known
for the corresponding integrals,
solutions can be obtained to arbitrary order in $\ep$.
This is different for the classes 5.2 and 5.3.
Here the pole structures of the differential equations
do not only contain poles of the form $1/x$ and $1/(1\pm x)$,
but in addition the poles $1/(1\pm 2x)$ (for 5.2) and $1/(1\pm x/2)$ (for 5.3).
It is not possible to integrate the differential equations
in terms of the usual harmonic polylogarithms with three weight functions,
Nevertheless we were able to integrate all equations
for these integrals up to order $\ep^{-1}$.
The $\ep^0$ parts of the master integrals 5.3 and 5.3a
were obtained analytically using Mellin-Barnes representation,
see below.

To get a result in higher orders in $\ep$
for the integrals of the classes 5.2 and 5.3 we
calculated the integrals in an expansion around $x=0$.
 With the help of the ansatz
\begin{equation}
\sum c_{ijk}\,\ep^i x^j \log^k x
\label{eq::ansatz}
\end{equation}
for the master integrals,
the differential equations can be expanded in $\ep$ and $x$.
As a result the differential equations reduce to algebraic equations
for the coefficients $c_{ijk}$.
In every order in $\ep$ there is one constant
$c_{ijk}$ which can not be determined with this procedure.

A problem which can arise when determining these remaining constants
is the fact that it may happen that the initial condition for some integrals
is fulfilled for the corresponding constant being arbitrary.
In the case of the class 5.3 this problem occurs
when one chooses the two integrals depicted in the third line
of Fig.~\ref{F:M5} as master integrals.
Sometimes one can determine the constants
via analytic considerations.
In our case this was not possible.
The problem can be solved by switching to another master integral basis
and replacing the integral with the dot with the integral 5.3b:
\newpage
\begin{eqnarray}
&& \frac{1}{(i\pi^{d/2})^3}
\int d^dl_1 d^dl_2 d^dl_3\\
&&\frac{\left(l_3-p\right)^2}%
{{\Big[m^2-l_2^2\Big] \left[m^2-\left(l_2-l_1\right)^2\right]
\left[M^2-\left(p-l_1\right)^2\right]^3 \Big[M^2-l_3^2\Big]
\left[M^2-\left(l_3-l_1\right)^2\right]}}\,,
\nonumber
\end{eqnarray}
where $p^2=M^2$ denotes the on-shell external momentum.
For this particular choice of master integrals
it is possible to determine all constants in the ansatz~(\ref{eq::ansatz})
using the corresponding initial conditions at $x=0$.

For the class 5.2 no basis of master integrals exists,
where all constants can be determined via the initial conditions.
In every order in $\ep$ the coefficient of the $x^3$-term in the
$x$-expansion of the integral 5.2 remains undetermined.
To get this constant we calculated the integral
by means of the method of regions~\cite{BS} up to the third order in $x$.
After having determined the constant in this way
all other orders in $x$ are fixed.

On the website mentioned earlier we present the analytic results
for the integrals depicted in Fig.~(\ref{F:M5}).
We also give expansions up to $\order{x^{14}}$ for the integrals
depicted in the second and third line of the figure up to $\order{\ep}$.

For the diagrams 5.1 and 5.1a in Fig.~\ref{F:M5},
we can obtain a one-fold Mellin--Barnes representation,
using~(\ref{S2:MB}):
\begin{eqnarray}
&&\raisebox{-9mm}{\begin{picture}(42,20)
\put(21,11.5){\makebox(0,0){\includegraphics{j51.eps}}}
\put(21,2){\makebox(0,0){$a_5$}}
\put(33,16){\makebox(0,0){$a_1$}}
\put(25,7){\makebox(0,0){$a_2$}}
\put(9,16){\makebox(0,0){$a_3$}}
\put(18,7){\makebox(0,0){$a_4$}}
\end{picture}} =
M^{2(d-a_3-a_4-a_5)} m^{d-2(a_1+a_2)}
\nonumber\\
&&{}\times
\frac{\Gamma(d/2-a_3) \Gamma(d/2-a_4) \Gamma(a_3+a_4-d/2)}%
{\Gamma(a_1) \Gamma(a_2) \Gamma(a_3) \Gamma(a_4) \Gamma(a_5)
\Gamma(d-a_3-a_4)}
\frac{1}{2\pi i} \int_{-i\infty}^{+i\infty} dz\,x^{-2z}
\nonumber\\
&&{}\times
\Gamma(-z) \Gamma(a_3+a_4+a_5-d-z) \Gamma(a_1+z) \Gamma(a_2+z)
\nonumber\\
&&{}\times
\frac{\Gamma(a_1+a_2-d/2+z) \Gamma(2(d-a_3-a_4)-a_5+2z)}%
{\Gamma(3d/2-a_3-a_4-a_5+z) \Gamma(a_1+a_2+2z)}\,.
\label{M5:MB1}
\end{eqnarray}

For the diagrams 5.2, 5.2a, 5.3 and 5.3a in Fig.~\ref{F:M5},
we have to use~(\ref{S2:MB}) twice.
Therefore, we obtain two-fold Mellin--Barnes representations:
\newpage
\begin{eqnarray}
&&\raisebox{-9mm}{\begin{picture}(42,20)
\put(21,11.5){\makebox(0,0){\includegraphics{j52.eps}}}
\put(21,2){\makebox(0,0){$a_5$}}
\put(33,16){\makebox(0,0){$a_1$}}
\put(25,7){\makebox(0,0){$a_2$}}
\put(9,16){\makebox(0,0){$a_3$}}
\put(18,7){\makebox(0,0){$a_4$}}
\end{picture}} =
\frac{M^{d-2a_5} m^{2(d-a_1-a_2-a_4-a_4)}}%
{\Gamma(a_1) \Gamma(a_2) \Gamma(a_3) \Gamma(a_4) \Gamma(a_5)}
\nonumber\\
&&{}\times \frac{1}{(2\pi i)^2}
\int_{-i\infty}^{+i\infty} dz_1 \int_{-i\infty}^{+i\infty} dz_2\,x^{-2(z_1+z_2)}
\nonumber\\
&&{}\times
\frac{\Gamma(-z_1) \Gamma(a_1+z_1) \Gamma(a_2+z_1) \Gamma(a_1+a_2-d/2+z_1)}%
{\Gamma(a_1+a_2+2z_1)}
\nonumber\\
&&{}\times
\frac{\Gamma(-z_2) \Gamma(a_3+z_2) \Gamma(a_4+z_2) \Gamma(a_3+a_4-d/2+z_2)}%
{\Gamma(a_3+a_4+2z_2)}
\nonumber\\
&&{}\times
\frac{\Gamma(a_5-d/2-z_1-z_2) \Gamma(d-a_5+2(z_1+z_2))}%
{\Gamma(d-a_5+z_1+z_2)}\,,
\label{M5:MB2}\\
&&\raisebox{-9mm}{\begin{picture}(42,20)
\put(21,11.5){\makebox(0,0){\includegraphics{j53.eps}}}
\put(21,2){\makebox(0,0){$a_5$}}
\put(33,16){\makebox(0,0){$a_1$}}
\put(25,7){\makebox(0,0){$a_2$}}
\put(9,16){\makebox(0,0){$a_3$}}
\put(18,7){\makebox(0,0){$a_4$}}
\end{picture}} =
\frac{M^{2(d-a_3-a_4-a_5)} m^{d-2(a_1+a_2)}}%
{\Gamma(a_1) \Gamma(a_2) \Gamma(a_3) \Gamma(a_4) \Gamma(a_5)}
\nonumber\\
&&{}\times \frac{1}{(2\pi i)^2}
\int_{-i\infty}^{+i\infty} dz_1 \int_{-i\infty}^{+i\infty} dz_2\,x^{-2z_1}
\nonumber\\
&&{}\times
\frac{\Gamma(-z_1) \Gamma(a_1+z_1) \Gamma(a_2+z_1) \Gamma(a_1+a_2-d/2+z_1)}%
{\Gamma(a_1+a_2+2z_1)}
\nonumber\\
&&{}\times
\frac{\Gamma(-z_2) \Gamma(a_3+z_2) \Gamma(a_4+z_2) \Gamma(a_3+a_4-d/2+z_2)}%
{\Gamma(a_3+a_4+2z_2)}
\nonumber\\
&&{}\times
\frac{\Gamma(a_5-d/2-z_1-z_2) \Gamma(d-a_5+2(z_1+z_2))}%
{\Gamma(d-a_5+z_1+z_2)}\,.
\label{M5:MB3}
\end{eqnarray}

We were able to get the $\ep^0$-part of the integrals 5.3 and 5.3a analytically
from~(\ref{M5:MB3}).
Since we have a reduction of this family to two master integrals we could
evaluate any pair of linearly independent integrals.
We have chosen Feynman integrals with the indices $(1,2,1,2,1)$
and $(1,1,1,1,2)$ (see the enumeration in Eq.~(\ref{M5:MB3})).
We evaluated it using (\ref{M5:MB3}) and resolving
singularities in $\ep$ in the corresponding two
MB representations~\cite{Smirnov:1999gc,Tausk:1999vh,books}
with the help of the \texttt{Mathematica} packages
\texttt{MB.m}~\cite{Czakon:2005rk}
and \texttt{MBresolve.m}~\cite{Smirnov:2009up}.
In both cases, to evaluate the $\ep^0$-part of these integrals,
we needed to evaluate at most two-fold finite MB integrals.
One of the integrations was done by corollary of Barnes lemmas
(see Appendix~D of~\cite{books}) implemented in \texttt{MB.m}.
At the last step, one-fold  MB integrals were evaluated by closing
the integration contour and summing up series.
The expansions in $x$ agree with those obtained from differential equations.
The results were also checked numerically using
the program \texttt{FIESTA}~\cite{Smirnov:2008py}
which implements sector decomposition.
where various algorithms of sector decomposition developed in
Refs.~\cite{BH,BW,Smirnov:2008py} are implemented.

For the calculations presented in~\cite{Bekavac:2007tk} we also needed
the $\ep^1$ coefficients of the five line master integrals. As discussed
above we were able to calculate all of them as series in $x$ but not all
could be evaluated analytically. Let us now describe how we can get a
numerical solution for these integrals, including the $\ep^1$
coefficient. We also used this method as a numerical
check of our analytical results.

We choose integral 5.3 (Fig.~\ref{F:M5}) with all indices equal to 1
as an example. 
The starting point is the MB representation, Eq. (\ref{M5:MB3}), of this
integral. Inserting $d=4-2\ep$ and $a_i=1$ we obtain
\begin{eqnarray}
&&\raisebox{-7.5mm}{\includegraphics{j53.eps}}
= \frac{M^{2-6\ep}}{(2\pi i)^2} \int_{-i\infty}^{+i\infty} dz_1
\int_{-i\infty}^{+i\infty} dz_2\,x^{-2\ep-2z_1}
\nonumber\\
&&\qquad{}\times
\frac{\Gamma(-z_1) \Gamma^2(1+z_1) \Gamma(\ep+z_1)
\Gamma(-z_2) \Gamma^2(1+z_2) \Gamma(\ep+z_2)}%
{\Gamma(2+2z_1) \Gamma(2+2z_2) \Gamma(3-2\ep+z_1+z_2)}
\nonumber\\
&&\qquad{}\times
\Gamma(-1+\ep-z_1-z_2) \Gamma(3-2\ep+2z_1+2z_2)\,.
\label{eq:I53}
\end{eqnarray}

The next step is the resolution of the $\ep$ singularities using the
algorithm described in~\cite{Smirnov:1999gc,Smirnov:2009up}. Our integral
is decomposed into five analytical expressions, two one-dimensional MB
integrals and one two-dimensional one, which contains the same integrand
as Eq.~(\ref{eq:I53}) but with modified integration contours.
The integrand expanded in $\ep$ reads
\begin{eqnarray}
F &=&- \frac{(x^2)^{-z_1}}{ \Gamma\left(z_1+\frac{3}{2}\right) 
\Gamma(z_1+z_2+3) \Gamma (2 z_2+2)}\\ \nonumber
&\times& 2^{2 z_2+1}\, \Gamma (-z_1)\, \Gamma (z_1)\,\Gamma (z_1+1)\,
\Gamma (-z_1-z_2-1)\,\Gamma (-z_2) \\ \nonumber
&\times& \Gamma (z_2)\, \Gamma^2(z_2+1)\,
\Gamma\left(z_1+z_2+\frac{3}{2}\right)\, \Gamma(z_1+z_2+2)\\ \nonumber
&\times& \Bigl\{ -1 + \ep \Bigl[ 2 \log x + 2 \log 2 -\psi(z_1)-\psi(-z_1-z_2-1)-\psi(z_2) \\ \nonumber
&+& \psi\left(z_1+z_2+\frac{3}{2}\right) +\psi(z_1+z_2+2)-2 \psi(z_1+z_2+3)\Bigr] \Bigr\}.
\end{eqnarray}

To evaluate the integral we close the integration contours and sum up
the residues. Since $x^2$ appears with the power $-z_1$ and we are
interested in an expansion for small $x$ we have to close the contour
for $z_1$ to the left. We are free in the way how we close the contour
for $z_2$ and choose the right side, considering the $\Gamma$ functions
$\Gamma(-z_2)$ and $\Gamma(1-z_1-z_2)$. The integral is then given as
the sum of the residues in the points $z_2=n$ and $z_2=-z_1+1+n$ for
positive integers $n$. The expressions for these residues read (we write
only the $\ep^0$ part here):
\begin{eqnarray}
&&A_1(n) = -\mathrm{Res}(F, z_2 = n)
= \frac{2^{2 n+1}\, (x^2)^{-z_1}}{n\, \pi\,(n+z_1+1)}
\nonumber\\
&&\times \frac{\Gamma^2(n+1)\, \Gamma (1-z_1)\, \Gamma (-n-z_1+1)\, 
                \Gamma (-z_1)\, \Gamma^2(z_1)\, \Gamma (z_1+1) }
          {\Gamma(2 n+2)\, \Gamma\left(z_1+\frac{3}{2}\right)\, 
            \Gamma(n+z_1+1)\, \Gamma (n+z_1+3)}
\nonumber \\
&&\times  \Gamma(n+z_1)\,\Gamma\left(n+z_1+\frac{3}{2}\right)\,
          \Gamma(n+z_1+2)\, \sin (\pi  (n+z_1)) \,,
\label{eq:A1}\\
&&A_2(n) = -\mathrm{Res}(F, z_2 = -z_1+1+n)
= \frac{-2^{2 n}\, (x^2)^{-z_1}}{2^{2 z_1+1}\,n\, (n-z_1-1)}
\nonumber \\
&&\times \frac{ \Gamma \left(n+\frac{1}{2}\right)\, \Gamma(n+1)\, \Gamma(1-z_1)\,
            \Gamma^2(n-z_1)\, \Gamma (-z_1)\, \Gamma^2(z_1)\, \Gamma(z_1+1)}
{ \Gamma (n) \Gamma (n+2) \Gamma (2 n-2 z_1) \Gamma \left(z_1+\frac{3}{2}\right)}\,.
\label{eq:A2}
\end{eqnarray}

Now we take the residues in the points $z_1=-n-\frac{3}{2}-m$ and
$z_1=-m$ with positive integers $m$. We have to choose $m \geq 0$ or $m
\geq 1$, depending on how the integration contour passes the poles of
the $\Gamma$ functions.

Thus we get a two-fold sum,
\begin{equation}
\label{eq:sumres}
\sum_n \sum_m f_i(m,n),
\end{equation}
the summands $f_i$ consisting of rational functions of $m,n$
and $\Gamma$, $\psi$ and trigonometric functions
whose arguments are linear combinations of $m,n$.

To check our analytical series expansion we proceed as follows.
Some of the summands are proportional to $(x^2)^{m+n}$
and some to $(x^2)^m$. In the first case we only need to sum up a finite
number of terms to get an expansion to a given order in $x$. In the
second case the sum is an expansion in $x$ where the coefficients are
infinite sums in $n$. We evaluate these coefficients to a finite order
given by the required numerical precision.

The one-fold sums contributing to the integral 5.3 are special cases of the
described procedure. However, there is one partial sum which does not
depend on $x$ and shows quite a bad convergence behaviour. In this case
we use the method of nonlinear sequence transformations (see
Refs. \cite{Weniger,Bekavac:2005xs} and references therein) to improve the
convergence. 

As a result we finally get an expansion for the integral 5.3 in powers of $x$
with numerical coefficients. Up to the order $x^5$ it reads 
\newpage
\begin{eqnarray}
&&\raisebox{-7.5mm}{\includegraphics{j53.eps}}
= M^{2-6\ep} e^{-3\gamma_E\ep} \biggl\{
\frac{1}{\ep^3} \bigl( -0.66667 - 0.33333 \,x^2 \bigr)
\nonumber\\
&&{} + \frac{1}{\ep^2} \Bigl[ -3.33333 + (-2.00000 + 2.00000 \log x) \,x^2\Bigr]
\nonumber\\
&&{} + \frac{1}{\ep} \Bigl[ -13.60147 + (-0.57606 + 12.00000 \log x -
  2.00000 \log^2 x) \,x^2
\nonumber\\
&&\qquad{} + (-5.03987 + 3.00000 \log x - 2.00000 \log^2 x) \,x^4 \Bigr]
\nonumber\\
&&{} + \Bigl[ -52.02282
\nonumber\\
&&\qquad{} + (31.35527 + 28.61586 \log x - 12.00000 \log^2 x + 1.33333 \log^3 x) \,x^2
\nonumber\\ 
&&\qquad{} - 52.63789 \,x^3
\nonumber\\
&&\qquad{} + (13.71171 - 0.92026 \log x - 5.00000 \log^2 x + 4.00000 \log^3 x) \,x^4
\nonumber\\
&&\qquad{} + 10.52758 \,x^5 \Bigr]
\nonumber\\
&&{} + \ep \Bigl[ -153.61196 + \bigl(149.09532 + 91.43326 \log x
\nonumber\\
&&\qquad{} - 28.61586 \log^2 x + 8.00000 \log^3 x - 0.66667 \log^4 x\bigr) \,x^2
\nonumber\\
&&\qquad{} + (-94.12475 + 210.55156 \log x) \,x^3
\nonumber\\
&&\qquad{} + \bigl(-106.98907 + 20.56599 \log x - 6.01454 \log^2 x + 6.00000 \log^3 x  
\nonumber\\
&&\qquad{} - 4.66667 \log^4 x\bigr) x^4
+ (-18.37249 - 42.11031 \log x) \,x^5 \Bigr] \biggr\}\,,
\label{eq:Summe}
\end{eqnarray}

where we arbitrarily display five digits. We determined the coefficients
with a relative error of at least $10^{-8}$. The result numerically
agrees with the analytical expansion obtained with the differential
equation method.

To find numerical solutions for the master integrals we insert 
fixed values for $x$ in the sum (\ref{eq:sumres}) and sum it up numerically. As
a further check we use the numerical integration routine of
\texttt{MB}. The summation procedure takes more computing time but its
advantage is that the errors are smaller than those we get from the
numerical integration.  

We used this method to find numerical expansions and also numerical
results for fixed values of $x$ for the master integrals 5.2, 5.2a, 5.3
and 5.3a. 
Let us finally mention the differences in the calculation of these.
Diagram 5.2 has two loops with the light quark mass. Therefore the MB
integrand is proportional to $x^{z_1+z_2}$ and we have no freedom to
choose how we close the integration contours. On the other hand, all
summands contain the parameter $x$ and thus the calculation is faster. 
Integrals 5.2a and 5.3a contain an additional power on one
denominator. This changes the arguments of the $\Gamma$ functions and we
have less MB integrals to consider. Their complexity is comparable to
that of the undotted integrals. 

\section{Master integrals with 6 lines}
\label{S:M6}

Contrary to the integrals with five lines,
the ones depicted in Fig.~\ref{F:M6} pose no difficulties
when solving them with the help of the differential equation method.
The integrals where some parts of the $\ep$-expansion are only known
as expansions in $x$ (see the previous Section) do not appear
as contracted classes in the differential equations.
Furthermore, the poles of the coefficients in the differential equations
are only of the form $1/x$ and $1/(1\pm x)$.
We can therefore get closed solutions of the equations
in terms of harmonic polylogarithms.

To obtain the integration constants we use the boundary condition at $x=1$.
The corresponding values can be found in~\cite{Melnikov:2000zc}
up to $\order{\ep}$.
We obtain for 6.1, 6.2, 6.3
\begin{eqnarray}
&&{}\hspace{-9mm}\raisebox{-7mm}{\includegraphics{j61.eps}} =
M^{-6\ep} \Gamma^3(1+\ep) \left(
\frac{1}{6 \ep^3} + \frac{3}{2 \ep^2}
- \frac{2 \pi^2 - 55}{6 \ep}
+ \order{1} \right)\,,
\nonumber\\
&&{}\hspace{-9mm}\raisebox{-7mm}{\includegraphics{j62.eps}} =
M^{-6\ep} \Gamma^3(1+\ep) \left(
\frac{1}{3 \ep^3} + \frac{7}{3 \ep^2}
+ \frac{31}{3 \ep}
+ \order{1} \right)\,,
\nonumber\\
&&{}\hspace{-9mm}\raisebox{-7mm}{\includegraphics{j63.eps}} =
M^{-6\ep} \Gamma^3(1+\ep) \left(
\frac{2 \zeta_3}{\ep}
+ \order{1} \right)\,.
\label{M6:1}
\end{eqnarray}
The results for the master integrals with six lines
up to $\order{\ep}$ can be obtained on the website.

We were able to derive a one-fold MB representation
for the master integral 6.1 in Fig.~\ref{F:M6}.
Using~(\ref{S2:MB}) we get a two-loop on-shell integral
with a single non-integer index $-z$.
It can be reduced to trivial ones by integration by parts,
and we obtain
\newpage
\begin{eqnarray}
&&\raisebox{-7mm}{\begin{picture}(42,17)
\put(21,8.5){\makebox(0,0){\includegraphics{j61.eps}}}
\put(33,13){\makebox(0,0){$a_1$}}
\put(24,4){\makebox(0,0){$a_2$}}
\end{picture}} =
\frac{M^{-4\ep} m^{4-2(a_1+a_2+\ep)}}{(1-2\ep) \Gamma(a_1) \Gamma(a_2)}
\frac{1}{2\pi i} \int_{-i\infty}^{+i\infty} dz\,x^{-2z}
\nonumber\\
&&\frac{\Gamma(-z) \Gamma(a_1+z) \Gamma(a_2+z) \Gamma(a_1+a_2-2+\ep+z)}%
{(1-2\ep+z) \Gamma(a_1+a_2+2z)}
\nonumber\\
&&\Biggl[ \Gamma(\ep)
\frac{\Gamma(\ep-z) \Gamma(2-2\ep+2z)}{\Gamma(2-2\ep+z)}
- \frac{\Gamma(\ep) \Gamma^2(1-\ep)}{\Gamma(1-2\ep)}
\frac{\Gamma(2\ep-z) \Gamma(2-4\ep+2z)}{\Gamma(2-3\ep+z)}
\nonumber\\
&&{} + \Gamma(1-\ep)
\frac{\Gamma(\ep-z) \Gamma(2\ep-z) \Gamma(1-\ep+z) \Gamma(2-4\ep+2z)}%
{\Gamma(-z) \Gamma(2-2\ep+z) \Gamma(2-3\ep+z)}
\Biggr]\,.
\end{eqnarray}

\section{Conclusion}
\label{S:Conc}

The status of our knowledge of the master integrals
is summarized in the Tables~\ref{T:S2}--\ref{T:M6}.
Here DE means that a result is known analytically, exactly in $x$,
and has been obtained by the method of differential equations;
MB --- the same, but using Mellin--Barnes representation;
$x$ --- expansion of the result (up to $x^{14}$) is known
analytically, from differential equations
(and a numerical value can be calculated for any $x$
by integrating the MB representation numerically);
a reference means that a result has been obtained by others.

\begin{table}[ht]
\caption{Two-loop sunset master integrals (Sect.~\ref{S:S2})}
\label{T:S2}
\begin{tabular}{c|cc}
&\includegraphics[scale=0.5]{s33.eps}
&\includegraphics[scale=0.5]{s31.eps}\\
&3.3&3.1, 3.1a\\
\hline
$\ep^{-2}$&DE&DE\\
$\cdots$&$\cdots$&$\cdots$\\
$\ep^3$&DE&DE\\
$\ep^4$&\cite{Broadhurst:1996az}&\cite{Argeri:2002wz}\\
$\ep^5$&\cite{Broadhurst:1996az}&\cite{Argeri:2002wz}\\
$\ep^6$&\cite{Broadhurst:1996az}&
\end{tabular}
\end{table}

\begin{table}[ht]
\caption{Three-loop vacuum and sunset master integrals
(Sects.~\ref{S:V3}, \ref{S:S3})}
\label{T:VS}
\begin{tabular}{c|ccc}
&\includegraphics[scale=0.5]{v2.eps}
&\includegraphics[scale=0.5]{v3.eps}
&\includegraphics[scale=0.5]{j48.eps}\\
&4.2&4.3, 4.3a&4.8, 4.8a, 4.8b\\
\hline
$\ep^{-3}$&DE&DE
&DE\\
$\cdots$&$\cdots$&$\cdots$&$\cdots$\\
$\ep^2$&DE&DE
&DE\\
$\ep^3$&\cite{Broadhurst:1996az}&&\cite{Mastrolia:2002tv}\\
$\ep^4$&\cite{Broadhurst:1996az}&&\\
$\ep^5$&\cite{Broadhurst:1996az}&&
\end{tabular}
\end{table}

\begin{table}[ht]
\caption{Three-loop master integrals with 5 lines (Sect.~\ref{S:M5})}
\label{T:M5}
\begin{tabular}{c|cccc}
&\includegraphics[scale=0.5]{j51.eps}
&\includegraphics[scale=0.5]{j52.eps}
&\includegraphics[scale=0.5]{j53.eps}
&\includegraphics[scale=0.5]{j54.eps}\\
&5.1, 5.1a&5.2, 5.2a&5.3, 5.3a&5.4, 5.4a\\
\hline
$\ep^{-3}$&DE&DE&DE&DE\\
$\ep^{-2}$&DE&DE&DE&DE\\
$\ep^{-1}$&DE&DE&DE&DE\\
$1$&DE&$x$&MB&DE\\
$\ep$&DE&$x$&$x$&DE\\
$\ep^2$&&&&DE
\end{tabular}
\end{table}

\begin{table}[ht]
\caption{Three-loop master integrals with 6 lines (Sect.~\ref{S:M6})}
\label{T:M6}
\begin{tabular}{c|ccc}
&\includegraphics[scale=0.5]{j61.eps}
&\includegraphics[scale=0.5]{j62.eps}
&\includegraphics[scale=0.5]{j63.eps}\\
&6.1&6.2&6.3\\
\hline
$\ep^{-3}$&DE&DE&DE\\
$\cdots$&$\cdots$&$\cdots$&$\cdots$\\
$\ep$&DE&DE&DE
\end{tabular}
\end{table}

These master integrals can be used for calculating any three-loop
diagrams with two legs of a massive particle (with mass $M$)
both of which are on the mass shell,
and there is a loop of another massive particle (with mass $m$).
They can be downloaded from~\cite{site}
in the form of \texttt{Mathematica} files.
The first physical application, namely the influence of $m_c\neq0$
on the on-shell mass of the $b$ quark and its wave-function
renormalization constant, has already been published~\cite{Bekavac:2007tk}.
We plan to use these results in some further problems.
A subset of master integrals considered here is also necessary
for another kind of problems --- diagrams with any number
of external lines having small momenta and containing loops of two
different massive particles.
There are several interesting physical problems which involve such diagrams.

We are grateful to M.~Steinhauser for collaboration~\cite{Bekavac:2007tk},
and to P.~Mastrolia and E.~Remiddi for providing the complete results
of Refs.~\cite{Argeri:2002wz,Mastrolia:2002tv}.

This work was supported by the Graduiertenkolleg
``Hochenergiephysik und Teilchenastrophysik''
and the Sonderforschungsbereich Transregio 9,
``Computergest\"utzte Theoretische Teilchenphysik''.
The work of V.S.\ was supported by the Russian Foundation for Basic
Research through grant 08-02-01451.
The work of D.S. was supported by the Alberta
Ingenuity foundation and NSERC.

\end{document}